\documentclass[prc,twocolumn,showpacs,floatfix,preprintnumbers,amsmath,amssymb,superscriptaddress]{revtex4-1}
\usepackage{amsfonts}

\usepackage{CJKutf8}
\usepackage{graphicx}
\usepackage{epsfig}
\usepackage{float}
\usepackage{dcolumn}  
\usepackage{bm}       
\usepackage{hyperref}
\usepackage{comment}
\newcommand{\hfbax}{\sc hfb-ax}

\newcommand{\hfbtho}{\sc hfbtho}
\newcommand{\hfodd}{\sc hfodd}
\newcommand{\madnesshfb}{\sc madness-hfb}
\newcommand{\madnesshf}{\sc madness-hf}
\newcommand{\madnesshfbcs}{\sc madness-hf+bcs}

\newcommand{\rr} {\boldsymbol{r}}
\newcommand{\bea}{\begin{eqnarray}}
\newcommand{\eea}{\end{eqnarray}}

\begin{document}

\begin{CJK*}{UTF8}{gbsn}

\title{Adaptive Multi-resolution 3D Hartree-Fock-Bogoliubov Solver for Nuclear Structure}

\author{J.C. Pei (裴俊琛)}
\affiliation{State Key Laboratory of Nuclear
Physics and Technology, School of Physics, Peking University,  Beijing 100871, China}
\affiliation{Department of Physics and Astronomy, University of
Tennessee, Knoxville, Tennessee 37996, USA}
\affiliation{Joint Institute for Nuclear Physics and Applications, Oak Ridge National Laboratory, Oak Ridge, Tennessee 37831, USA}

\author{G.I. Fann}
\affiliation{Computer Science and Mathematics Division, Oak Ridge
National Laboratory, Oak Ridge, Tennessee 37831, USA}

\author{R. J. Harrison}
\affiliation{Institute for Advanced Computational Science, Stony Brook University, Stony Brook, New York 11794, USA}
\affiliation{Computational Science Center, Brookhaven National Laboratory, Upton, New York  11973, USA}

\author{W. Nazarewicz}
\affiliation{Department of Physics and Astronomy, University of
Tennessee, Knoxville, Tennessee 37996, USA}
\affiliation{Physics Division, Oak Ridge National Laboratory, Oak Ridge, Tennessee 37831, USA}
\affiliation{Institute of Theoretical Physics, Faculty of Physics, University of Warsaw, ul. Ho{\.z}a 69, PL-00681 Warsaw, Poland}

\author{Yue Shi (石跃)}
\affiliation{Department of Physics and Astronomy, University of Tennessee, Knoxville, Tennessee 37996, USA}
\affiliation{Joint Institute for Nuclear Physics and Applications, Oak Ridge National Laboratory, Oak Ridge, Tennessee 37831, USA}

\author{S. Thornton}
\affiliation{Institute for Advanced Computational Science, Stony Brook University, Stony Brook, New York 11794, USA}

\begin{abstract}
\begin{description}
\item[Background] Complex many-body systems, such as triaxial and reflection-asymmetric nuclei, weakly-bound halo states, cluster configurations,  nuclear fragments produced in heavy-ion fusion reactions, cold Fermi gases, and pasta phases in neutron star crust, they are all characterized by large sizes and  complex topologies, in which many geometrical symmetries characteristic of ground-state configurations are broken. A tool of choice to study such complex forms of matter is an adaptive multi-resolution wavelet analysis. This method has generated much excitement since it provides a
common framework linking  many diversified methodologies across
different fields, including signal processing, data compression,
harmonic analysis and operator theory, fractals, and quantum field theory.

\item[Purpose]  To describe complex superfluid many-fermion systems, we
introduce an adaptive pseudo-spectral method for solving
self-consistent equations of nuclear
density functional theory in three dimensions, without symmetry restrictions.

\item[Methods] The numerical method is based on the multi-resolution
  and computational harmonic analysis techniques with multiwavelet
  basis.  The application of state-of-the-art in parallel programming
  techniques include sophisticated object oriented templates which parses the
  high-level code into distributed parallel tasks with a multithread
  task queue scheduler for each multicore node.  The inter-node
  communications are asynchronous.  The algorithm is variational and
  is capable of solving coupled complex-geometric systems of equations
  adaptively, with functional and boundary constraints, in a finite
  spatial domain of very large sizes, limited by existing parallel
  computer memory. For smooth functions, user defined finite
  precision is guaranteed.

\item[Results] The  new adaptive multi-resolution Hartree-Fock-Bogoliubov (HFB) solver {\madnesshfb} is benchmarked against a two-dimensional    coordinate-space solver {\hfbax} based on B-spline technique and three-dimensional solver {\hfodd}
based on the harmonic oscillator basis expansion. Several examples  are considered, including
self-consistent
HFB problem   for spin-polarized trapped cold fermions and
Skyrme-Hartree-Fock (+BCS) problem for triaxial deformed nuclei.

\item[Conclusions] The new {\madnesshfb} framework  has many attractive features when applied to nuclear and atomic  problems involving many-particle superfluid systems. Of particular interest are
weakly-bound   nuclear configurations  close to particle drip lines, strongly elongated and dinuclear configurations such as those present in fission and heavy ion fusion, and exotic  pasta phases that appear in the neutron star crust.
\end{description}
\end{abstract}

\pacs{21.60.Jz,31.15.E-,03.65.Ge,07.05.Tp,67.85.-d,03.75.Hh}

\maketitle
\end{CJK*}

\section{Introduction}

The roadmap for nuclear structure theory includes QCD-derived (or inspired)
nuclear interactions, ab-initio calculations for light and medium nuclei,
configuration interaction approaches for near-magic systems, and density
functional theory and its extensions for heavy, complex nuclei~\cite{(Bog13)}.
On the  road to the quantitative understanding of nuclear structure and
reactions, high-performance computing plays an increasingly important role.  As
stated in the recent decadal survey of nuclear physics \cite{Decadal2012}
``High performance computing provides answers to questions that neither
experiment nor analytic theory can address; hence, it becomes a third leg
supporting the field of nuclear physics."  Largest collaborations in
computational nuclear structure and reactions involve nuclear theorists,
computer scientists, and applied mathematicians to break analytic, algorithmic,
and computational barriers \cite{(Bog13),UNEDF}. This paper offers an example
of such a joint collaborative effort in the area of  nuclear Density Functional
Theory (DFT).

A key element of any  DFT framework is a HFB solver that computes
self-consistent solutions of HFB (or Bogoliubov-de Gennes) equations.
Traditionally, the HFB solvers in nuclear physics are based on the basis
expansion method, usually  employing harmonic oscillator wave functions
\cite{(Sto05),(Sto13),(Dob04b),(Sch12)}. These methods are very efficient but
they require huge bases for cases involving weakly-bound systems and large
deformations~\cite{(Mic08),(Kor12)}.  On the other hand, solving HFB equations
directly in coordinate-space can offer very precise
results~\cite{(Dob84),(Dob96),(Pei08)}. Unfortunately, current HFB calculations
for non-spherical geometries are computationally challenging.  There exist 2D
coordinate-space HFB solvers, based on B-splines, which have provided precise
descriptions of describing weakly bound nuclei and large
deformations\cite{(Pei08),(Ter03)}.  However, the extension from 2D to fully 3D
HFB calculations adds at least three orders of computational complexities (for
some recent developments, see Refs.~\cite{(Ste11),(Has13),(Bul13)}).

Similar to Fourier analysis, wavelet analysis deals with expansion of functions
in terms of basis functions.  Unlike Fourier analysis, wavelet analysis expands
functions not in terms of trigonometric functions but in terms of wavelets,
which are generated by translations and dilations of a fixed function, called
the mother wavelet.  The wavelets obtained in this way have special scaling
properties.  They are localized in time and frequencies, permitting more
precise local connection between their coefficients and the function being
represented.  These estimates allow greater numerical stability in
reconstruction and manipulation with controlled precision and sparsity.  For
example, the JPEG2000 compression algorithms were built using wavelets.  The
decoding could be accomplished in multiple ways and enabled scalable
compression with different resolution representations.  By truncating the data
stream early, a lower resolution image representation is obtained.

Multiwavelets consist of a set of wavelets.  The Alpert multiwavelets
\cite{(Alp93)} that we use are constructed from Legendre polynomials.  They are
discontinuous and singular orthonormal functions which permit better
approximations of singular and discontinuous functions with reduced Gibbs
effects.  Another feature is the availability of high vanishing moments, which
permit the sparse representation and application of smooth functions and many
singular operators, in finite precision.  Families of multiwavelets permit high
orders of approximations with fewer levels of refinement, which is essential
for efficient scientific computing.

To this end, we have developed a 3D {\madnesshfb} solver for HFB equations and
Hartree-Fock (HF) equations, which is a multi-resolution, adaptive spectral approximations
based solver, using a multiwavelet basis, with a scalable parallel
implementation~\cite{madness}.  The new framework is applied to polarized
ultracold Fermi gases in elongated optical traps as well as triaxial nuclei. In
both cases, we will demonstrate the capability of very large box calculations
which is essential for descriptions of complex geometries and topologies.

This paper is organized as follows. Section~\ref{madness} briefly introduce the
multiresolution mathematics, low-separation rank approximation, and parallel
runtime environment.  The iterative algorithm applied in {\madnesshfb}  is
presented in Sec.~\ref{strategy}.  In Sec.~\ref{benchmarks}, we benchmark
{\madnesshfb} solutions for cold fermions and nuclei.  Finally, conclusions are
given in Sec.~\ref{conclusions}.

\section{MADNESS-HFB framework}\label{madness}

Our implementation of {\madnesshfb}  uses the Multiresolution Adaptive
Numerical Environment for Scientific Simulations (MADNESS)
framework~\cite{madness}.  MADNESS is based on computational harmonic analysis
and nonlinear approximations using Alpert's multiwavelet basis
\cite{(Mal89),(Alp93),(Alp02)} to represent functions.  Fast parallel code
development and scalable performance have been possible due to the ease of
programming based on object-oriented abstractions for interprocessor
communications, multithreading and mathematical operations.

\subsection{Mathematics of MADNESS}

The mathematics implemented in the MADNESS software are based on
multiresolution analysis (MRA) \cite{(Mal89),(Alp02)}, nonlinear
approximations, and pseudo-spectral techniques.  There are two types of
techniques used in MADNESS to approximate functions and operators.  The first
is the use of multiresolution analysis based on Alpert's multiwavelets
\cite{(Alp93)}.  The second technique is the use of the low-separation rank
approximations of Green's functions based on Gaussian functions
\cite{(Bey02),(Har04)}.  In the following, we follow the notation and
derivations of Ref.~\cite{(Alp02)}.

\subsubsection{Multiresolution analysis with wavelets}\label{multires}

The application of MRA separates the behavior of functions and operators at
different length scales in a systematic expansion.  A consequence of the
separation of scales is that each operator and wave function has a naturally
independent adaptive refinement structure, reflected in terms of significant
expansion coefficients of desired precision.  The thresholding and truncation
of expansion coefficients below a user-defined error provides adaptive blocks
of non-trivial coefficients for a pseudo-spectral expansion.  The union of the
domains of the multiwavelets with non-zero coefficients provide an adaptive
dyadic spatial localization of the relevent contributions for the corresponding
refinement levels.  In 1D, the non-zero sets define an adaptive dyadic
refinement and correspondingly in 3D a pruned octtree type refinement.

The MRA representation used in MADNESS is analogous to that used in an adaptive
hp-SEM (spectral element method), which employs elements of variable size $h$
and piecewise-polynomial approximations of degree $p$.  By suitably refining
the mesh through $h$-refinements (dividing the volume elements into smaller
pieces) and $p$-refinements (increasing the polynomial degree in the expansion
within the elements) one can reach exponential convergence~\cite{(Bab92)}.  In
MADNESS, for each function or operator, the union of the domains of the
multiwavelet basis functions with non-zero coefficients, after thresholding,
defines an adaptive and heirarchical $h$-structure and the associated
multiwavelets form the set of the piecewise polynomials up to order $p$.  Thus,
there are multiple $h-p$ refinement structures that are used simultaneously.

The basis of scaling functions in 1D  is constructed in terms of the  normalized
Legendre polynomials rescaled to the unit interval $(0,1)$ and zero elsewhere.
For each level $n$ (defining the  volume refinement), the rescaled and translated basis function is given by:
\begin{equation}\label{wavelet}
{\cal \phi}_{il}^{n} (x)= 2^{n/2} {\cal \phi}_{i} ( 2^{n} x- l ),
\end{equation}
where ${\cal \phi}_i (x) = \sqrt{2i+1} P_i (2x-1)$, with $P_{i}(x)$ being the Legendre
polynomial on $(-1,1)$, and is $0$ elsewhere for $l=0, ..., 2^{n}-1$.
The basis functions (\ref{wavelet}) at level $n$ have domain of width $2^{-n}$.

Let $V_{n} = \{ {\cal \phi}_{il}^{n} (x), i=0,
..., k-1 \}$ be the span of the subspace at level $n$.  Let $W_{0}
= \{ {\cal \psi}_{i}(x) \}$ denote an orthonormal basis which
spans the difference subspace $V_{1} - V_{0}$.  These functions are called
multiwavelets.  As with the scaling functions let ${\cal
  \psi}_{il}^{n} (x)$ and $W_{n}$ denote the rescaled and shifted
multiwavelets and the corresponding subspace spanned by
these functions at level $n$.
The definition of scaling functions and
multiwavelets defines an ascending sequence of subspaces
\begin{equation}
V_{0} \subset V_{1} \subset V_{2} ...\subset V_{n}
\end{equation}
and
\begin{equation}
V_{n}  = V_{0} \oplus W_{0} \oplus ... \oplus W_{n-1},
\end{equation}
where the $\oplus$ denotes orthogonal sum.  The dimension of $V_{i}$
is greater than dimension of the subspace $V_{i-1}$; thus, the basis
functions of $V_{i-1}$ and $W_{i-1}$ can be written exactly in terms
of the basis functions of $V_{i}$.  These heirarchical linear algebraic relations
between the bases defines the 2-scale refinement structure between the
coefficients at level $i-1$ and $i$,
and fundamentally defines the adaptive structure with a given threshold truncation.

A smooth function $f(x)$ in the subspace $V_{n}$ can be approximated in terms of scaling functions as:
\begin{equation}
f(x) = \sum_{l=0}^{2^{n}-1} \sum_{j=0}^{k-1} s_{jl}^{n} {\cal \phi}_{jl}^{n}(x).
\end{equation}
Represented in the multiwavelet basis, $f(x)$ is
\begin{equation}
f(x) = \sum_{j=0}^{k-1} s_j {\cal \phi}_{j}(x) + \sum_{j=0}^{k-1} \sum_{m=1}^{n-1} \sum_{l=0}^{2^{m}-1} d_{jl}^{m} {\cal \psi}_{jl}^{m} (x),
\end{equation}
with $s_{jl}^{n} = \int_{2^{-n} l}^{2^{-n}(l+1) } f(x) {\cal \phi}_{jl}^{n} dx$
and $d_{jl}^{m} = \int_{2^{-m} l}^{2^{-m}(l+1) } f(x) {\cal \psi}_{jl}^{m} dx$.

\begin{figure}[h!]
\includegraphics[width=\columnwidth]{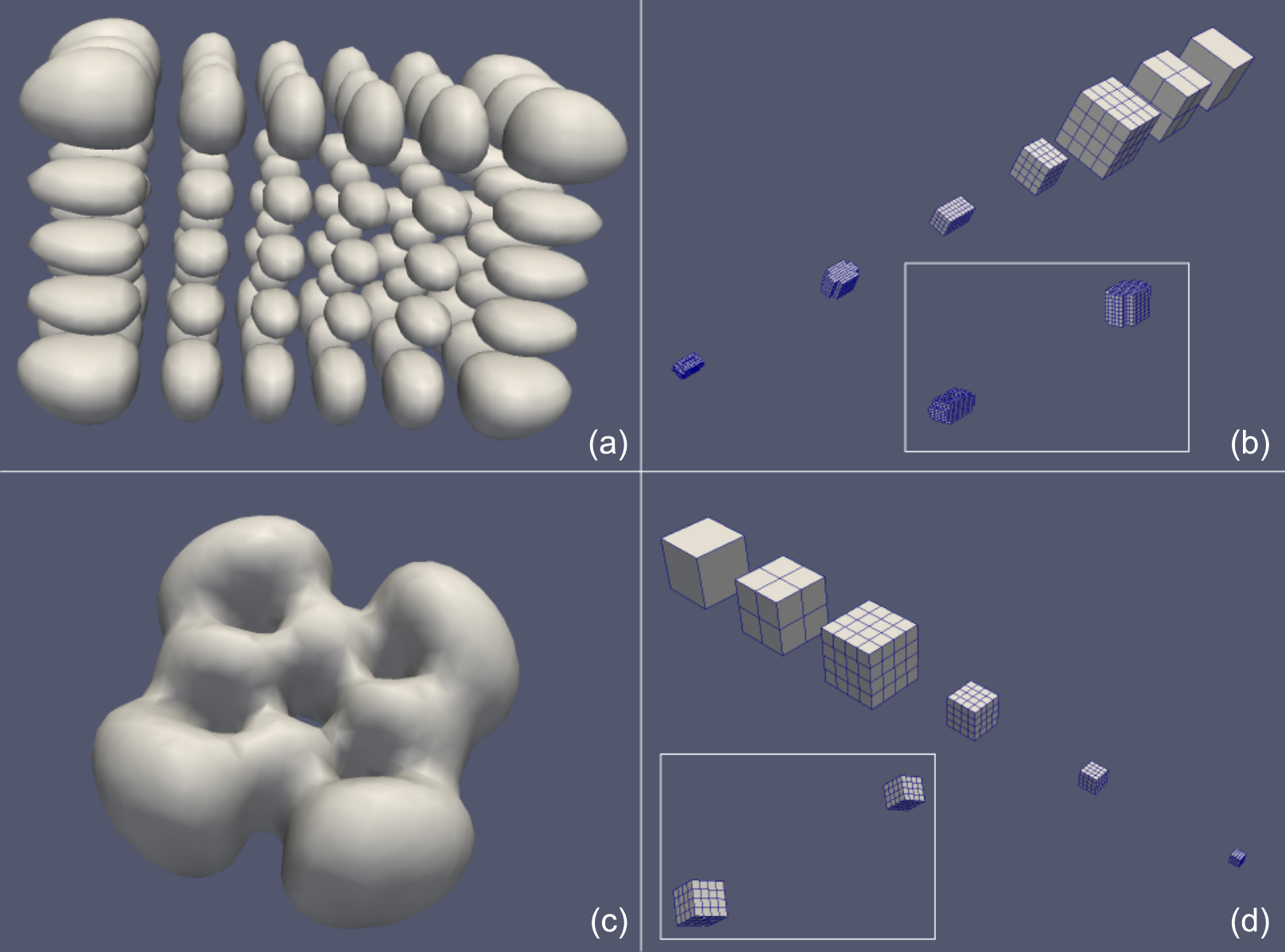}
\caption{Pedagogical illustration of adaptive representations in
MADNESS-HFB. Top: (a) the  modulus squared of the single-neutron wave function corresponding to the single-particle energy of $-5.214$\,MeV obtained in 
{\madnesshf} calculations for $^{110}$Mo 
(see Sec.~\ref{HFBCS} for details), and (b) the corresponding spectral refinement
structure. Bottom: (c) the  modulus squared of the  single-proton wave function corresponding to the energy eigenvalue $-12.272$\,MeV in $^{110}$Mo and its adaptive spectral structure (d).
Notice that the refinement structure 
for the proton wave function is similar to a truncated octtree type of refinement
but the structure for neutron wave function is more complicated,
especially at the finer level, see insets in panels (b) and (d).
\label{mr3dhfb}
}
\end{figure}

In the discussion above, we described the representations based on
multiwavelets in 1D.  In 3D applications,  we use tensor products of 1D
multiwavelets as well as scaling functions in non-standard form.
Figure~\ref{mr3dhfb} illustrates  the multiresolution structure of sample
wave functions.  

For smooth functions the computational methodologies are guaranteed to
approximate the solutions to the desired user precision $\epsilon$, with
respect to the relative norm, with the correct number of digits specified by
the error.  The estimate is based on truncating the difference coefficients in the multiwavelet expansion, 
\begin{equation}
||d^n_l||_2=\sqrt{\sum_{j}|d_{jl}^n|^2 } \le \epsilon~{\min}(1,2^{-n} L),
\label{truncat}
\end{equation}
where $L$ is the minimum of the width of the computational domain.

\subsubsection{Multiresolution}\label{MRA-HB}

For the one-body Schr\"odinger equation,
\begin{equation}\label{SE}
\left(-\Delta  +V \right)\psi = E\psi,
\end{equation}
 the usual diagonalization approach is also derived and used.  In this case, given
a basis $\psi_{i}$, a Hamiltonian matrix is formed
\begin{equation}
H_{i,j} = \langle\psi_i | -\Delta +V | \psi_j \rangle;\ S_{i,j} = \langle\psi_i |\psi_j \rangle,
\end{equation}
to form a generalized eigenproblem
$H \psi = S \psi.$

A generalized eigensolver computes the eigenvalues and the eigenvectors.  The
eigenvectors are coefficients with respect to the multiwavelets basis, and they
are converted back to the spectral representation for further computation. The
Laplacian $\Delta$, the potential $V$, and the wave-functions $\psi_i$ are all
in MRA form. The derivatives of multiwavelets are expanded in terms of
multiwavelets, and the coefficients are tabulated.  By linearity, the
derivatives of a function can be computed by matrix-vector products, or
tensor-tensor products in higher dimensions, using only the multiwavelet
coefficients.

This procedure permits computation of ``self-consistent'' solutions of DFT equations.

\subsubsection{Low-separation rank approximation of Green's functions}\label{LSR}
Recall that the (one-body) Schr\"odinger equation (\ref{SE})
can be rewritten as a Lippmann-Schwinger equation as
\begin{equation}\label{LS}
(\Delta + E) \psi = V \psi.
\end{equation}
There are several advantages of using the integral form (\ref{LS})
over the differential form (\ref{SE}). Namely, the integral form
provides higher accuracy as high-frequency noise is attenuated not
amplified, builds correct asymptotics, good condition number, and is
potentially more computationally efficient. In most bases Green's
function representation is often dense, and the use of multiresolution
analysis and multiwavelets provides fast algorithms with sparse
structure in finite floating arithmetic with guaranteed precision.  If
no controlled truncations of the multiwavelet coefficients are performed the representation of the Green's function and its application will be dense.

The formal solution of (\ref{LS}) can be written as:
\begin{equation}
\psi(r) = \int_{-\infty}^{\infty} G(r-r') V \psi(r')  dr' = (G \star V\psi),
\end{equation}
where the Green's function $G(r)$ is the Helmholtz kernel and the symbol $\star$ represents convolution.
If the eigenvalue is bound ($E < 0$), the Green's function is the Yukawa potential $\exp(-kr)/r$ where $k=\sqrt{- E }$.  In general, one works with $G=(\Delta + E+ i\varepsilon )^{-1}$ with $ \varepsilon  \rightarrow +0$ and specifies how to integrate around the poles.

For bound-states, a low-separation rank (LSR) expansion \cite{(Bey02),(Har04)} of the Yukawa potential is used,
\begin{equation}
\frac{e^{-kr}}{r} = \sum_{l} {\sigma}_{l} e^{-{\tau}_{l} r^2} + O(\epsilon).
\end{equation}
The LSR approximation represents Green's function in terms of a Gaussian expansion.
Such a form reduces the application of 3D convolutions
to an set of uncoupled 1D convolutions with the number of terms
scaling logarithmically with respect to the relative precision $\epsilon$.  Since the convolution operator is linear, tables of
precomputed transformation matrices with respect to the multiwavelets
enable fast applications of  convolutions~\cite{(Fan05)}.

The technique described above to solve the Schr\"odinger equation can be
directly applied to a HF problem, and -- after a minor generalization -- to HFB
equations.

\subsection{MADNESS parallel runtime environment}

A novel parallel execution runtime environment has been implemented in the
MADNESS software library.  MADNESS uses one Message Passing Interface process
to communicate between nodes, and  POSIX Threads within a node to exploit
shared memory parallelism with a global addressable view of memory space in
software. The MADNESS runtime is based on a parallel task-based computing model
with a graph-based scheduler and a task queue on each node, to enable
distributed multi-threaded computation.  A microparser is used to decouple
tasks as much as possible but also to detect data dependencies so as many
independent and out-of-order tasks can execute simultaneously, ensuring correct
and minimal number of synchronization and thread termination.

Although the dedicated use of a core for inter-node communication and a core
for handling thread scheduling may be a big sacrifice of computational
resources, for supercomputers with large numbers of cores per node, we are able
to obtain more than 50\% of peak core performance for the remaining cores.
Most scientific and engineering codes obtain only about 10\% of the peak
processor performance.

The flexibility of {\madnesshfb} in its design and programming style permits
the solution of multiphysics problems with complex geometric structures and
boundary conditions in large volumes in the coordinate-space formulation --
limited only by the size of aggregate computer memory.  Nuclear fission, exotic
topologies in super- and hyperheavy nuclei, neutron star crusts, and cold atoms
in elongated traps are some examples which can take advantage of these
features.

\section{{\sc\bf madness-hfb} Strategy}\label{strategy}

The general HFB equation for a two-component (e.g., spin-up $\uparrow$ and spin-down $\downarrow$) system of fermions can be written as \cite{(Bul08),(Bul08a),(Ber09),(Pei10)}:
\begin{equation}
  \left[
\begin{array}{cc}%
 h_\uparrow-\lambda_{\uparrow} & \Delta \\
 \Delta^* & -h_\downarrow +\lambda_\downarrow
\end{array}
\right]\left[
\begin{array}{c}%
u_i \\
v_i
\end{array}
\right]=E_i\left[
\begin{array}{c}%
u_i \\
v_i
\end{array}
\right],
\label{BdG}
\end{equation}
where  $h_\uparrow$ and $h_\downarrow$ are the
Hartree-Fock Hamiltonians for the spin-up and spin-down components, respectively.  The corresponding chemical potentials are denoted as $\lambda_\uparrow$ and $\lambda_\downarrow$, and  the pairing potential is $\Delta$.

There are two standard approaches to solve the HFB equation (\ref{BdG}). In the
basis expansion method, eigenvectors $(u_i,v_i)$ are expressed in terms of a
single-particle basis and the self-consistent procedure applies the HFB
Hamiltonian matrix diagonalization. The HFB solvers  {\hfbtho} \cite{(Sto05)}
(using the cylindrical transformed deformed harmonic-oscillator basis) and
{\hfodd} \cite{(Sch12)} (using the Cartesian deformed harmonic-oscillator
basis), employed in this work to benchmark {\madnesshfb} belong to this class.
A second way is to solve the HFB equations in the coordinate-space by finite
difference or finite element methods \cite{(Ter03),(Ben05b),(Pei08)}  or  in
the momentum space using fast Fourier transform~\cite{(Bul08b)}.  The  strategy
applied in {\madnesshfb}, described in Sec.~\ref{madness},  combines  features
from these two  approaches.  The original method has been developed in the
context of HF and DFT problems in computational
chemistry~\cite{(Har04),(Fan09)}.

To illustrate the self-consistent procedure, let us consider a case
of an unpolarized system ($h_\uparrow$=$h_\downarrow$) with constant effective mass $1/\alpha$.
The mean-field Hamiltonian  is
\begin{equation}
h(\rr)=-\frac{\alpha \nabla^2}{2}+U(\rr),
\end{equation}
where  $U(\rr)$ is the HF potential.
As discussed in Sec.~\ref{LSR}, it is convenient to rewrite the HFB equation
(\ref{BdG})
in a Lippmann-Schwinger form. To this end,  in each step of HFB iterations, we introduce the Green's functions $G_+$ and $G_-$:
\begin{equation}\label{Greenf}
G^n_\pm = \frac{1}{\pm {\alpha \nabla^2 \over 2}+(E_s \pm E_i^n)},
\end{equation}
where  $E^n_i$ is the $i$-th HFB eigenvalue in the $n$-th iteration step, and $E_s$ is the energy displacement that shifts  the positive-energy HFB  eigenvalues so that the associated Green's function is properly defined.

To solve the self-consistent HFB eigenproblem, the HFB wave functions can be updated by as follows,
\begin{subequations}
\begin{eqnarray}\label{conv}
u_i^{n+1} & =  \left(G_+ \star \left[(U-\lambda)u_i^n+\Delta v_i^n+ E_s u_i^n \right]\right)\\
v_i^{n+1} & = \left(G_-\star \left[(U-\lambda)v_i^n-\Delta u_i^n+ E_s v_i^n\right]\right)
\end{eqnarray}
\end{subequations}
Following this strategy, in the following section, we use {\madnesshfb} to solve  HFB problems with
advanced local energy density functionals for cold fermions and nuclei.

\section{Benchmark Problems}\label{benchmarks}

In this section, the {\madnesshfb} framework is benchmarked   by solving  HFB equations  for the trapped unitary Fermi gas and HF-BCS equations  for a triaxial nucleus.  The {\madnesshfb} solutions for atoms and nuclei are compared with results of  2D
{\hfbax} and 3D {\hfodd} calculations, respectively.

\subsection{HFB solver for unitary Fermi gas}

The unitary limit of Fermi gas, is characterized by an infinite $s$-wave
scattering length. Of particular interest are superfluid phases in
spin-imbalanced systems, such as the Fulde-Ferrell-Larkin-Ovchinnikov
\cite{(Ful64),(Lar65)} phase that exhibits oscillated pairing gaps.  The
ultracold Fermions at the unitary limit can be described by the superfluid
density functional SLDA~\cite{(Bul07)} and its asymmetric extension ASLDA for
spin-polarized systems~\cite{(Bul08)}.

The single-particle Hamiltonian of the ASLDA  for asymmetric systems can be written as ~\cite{(Bul08)}:
\begin{equation}
h_\sigma=-\frac{\hbar^2}{2m}\boldsymbol{\nabla}\cdot(\boldsymbol{\nabla}\alpha_\sigma(\rr))+U_\sigma(\rr)+V_{\rm ext}(\rr),
\end{equation}
where $\sigma=(\uparrow,\downarrow)$ denotes the spin up and spin down components.
The local polarization is denoted as $x(\rr)=\rho_\downarrow(\rr)/\rho_\uparrow(\rr)$ with $ x(\rr) \leqslant 1$,
where $\rho_\uparrow(\rr)$ and $\rho_\downarrow(\rr)$ are densities of spin-up and spin-down atoms, respectively.
The total polarization of the system is $P=(N_\uparrow-N_\downarrow)/N$. The quantity
$\alpha_{\sigma}(x(\rr))$ is
the local effective mass.
The SLDA formalism can be obtained from ASLDA by assuming $x(\rr)=1$, resulting in identical effective masses and Hartree potentials for spin-up and spin-down species.

The cold atoms are trapped in an external potential
\begin{equation}
V_{\rm ext}(x,y,z) = V_0\left[1-e^{-\frac{\omega^2(x^2+y^2+z^2/\eta^2)}{2V_0}}\right],
\end{equation}
where the trap aspect ratio $\eta$ denotes the elongation of the optical trap potential.
The equations are normalized so that $\hbar=m=\omega=1$ (trap units). All other details pertaining to our SLDA and ASLDA calculations closely follow Ref.~\cite{(Pei10)}.

We first consider an  SLDA case of  ten particles in a spherical trap with
$V_0=10$ and the quasiparticle energy  cutoff $E_{\rm cut}=6$.  The
calculations were performed in a 3D box  $(-L, L)^3$ with $L=60$ With this box
and cutoff, the self-consistent HFB solution involves 296  one-quasiparticle
eigenfunctions.
In the present SLDA and ASLDA benchmark calculations, we adopt wavelet order of
$p=8$ with a requested truncation precision of $\epsilon=10^{-5}$ (see Eq.\ref{truncat}).

The {\madnesshfb} results were benchmarked using the 2D HFB solver {\hfbax}.
In the {\hfbax} calculation, the maximum mesh size is $0.3$, the order of
B-splines is $k=12$, and the box size is $R_{max}=Z_{max}=14$.  The eigenvalues
and occupation numbers of some of the lowest and highest states from the two
codes are compared in Table \ref{tabho1}. The agreement is excellent, also for
the total energy and chemical potential.

\begin{table}[htb]
 \caption{\label{tabho1}
 Benchmark comparison of  {\madnesshfb} and {\hfbax} results for $10$ particles in the spherical trap without polarization. Displayed are:
  one-quasiparticle energies $E_i$, occupations $v_i^2$,  chemical potential
   $\lambda$, and total energy $E_t$. Each  one-quasiparticle  state is labelled by means of orbital quantum number $\ell$ and parity $\pi=(-1)^\ell$. Note that every solution is $2\ell+1$-folded degenerate with respect to the magnetic quantum number.
The numbers in parentheses denote powers of 10. The energy is expressed in
trap units ($\hbar$=$m$=$\omega$=1).}
\begin{ruledtabular}
\begin{tabular}{cccccc}
  & & \multicolumn{2}{c}{\madnesshfb}  &  \multicolumn{2}{c}{\hfbax} \\
$i$ & $\ell$  &  $E_i$  &  $v_i^2$  &  $E_i$  &   $v_i^2$ \\
\hline
\\[-7pt]
1 & 0 &  0.90394 &  0.23240  &   0.9039{\bf 5} &  0.23240\\
2 & 2 &  1.06340  &  0.17779 &  1.0634{\bf 2}  &  0.17779\\
3 & 1 &  1.12686 &  0.47471  &  1.1268{\bf 8}  &  0.4746{\bf 9}\\
4 & 3 &  1.92205 &  2.2491(-2)  &  1.9220{\bf 6}  &  2.2491(-2)\\
5 & 1 &  2.00891 &  0.28448 &  2.0089{\bf 4}  &  0.2844{\bf 9} \\
6 & 0 &  2.54095 &  0.30390 &  2.5409{\bf 6}  &  0.3039{\bf 3} \\
7 & 2 &  2.69803 &  3.3837(-2)  &  2.6980{\bf 4}  &  3.383{\bf 8}(-2)\\
8 & 0 &  2.82496 &  0.60883 &  2.8250{\bf 0}  &  0.6088{\bf 4}  \\
9 & 4 &  2.91835 &  3.8699(-3)  &  2.9183{\bf 6}  &  3.869{\bf 8}(-3)\\
10 & 1 &   3.44774 & 2.3162(-2) &  3.4477{\bf 5}  &  2.316{\bf 5}(-2)\\
21 & 7 &  5.54071 &  3.1957(-5)  &  5.5407{\bf 2}  &  3.195{\bf 4}(-5)\\
22 & 2 &  5.58728 & 3.6548(-3) &  5.58728  &  3.655{\bf 0}(-3)\\
23 & 4 &  5.75254 & 1.8024(-3)  &  5.7525{\bf 5}  &  1.8024(-3)\\
\hline
\\[-7pt]
   &  & \multicolumn{2}{c}{$E_t =18.5641$} &  \multicolumn{2}{c}{$E_t=18.563{\bf 9}$} \\
  &  & \multicolumn{2}{c}{ $\lambda=2.24917$}  &  \multicolumn{2}{c}{$\lambda=2.2491{\bf 6}$}
\end{tabular}
\end{ruledtabular}
\end{table}

\begin{figure}[t]
 \includegraphics[width=0.85\columnwidth]{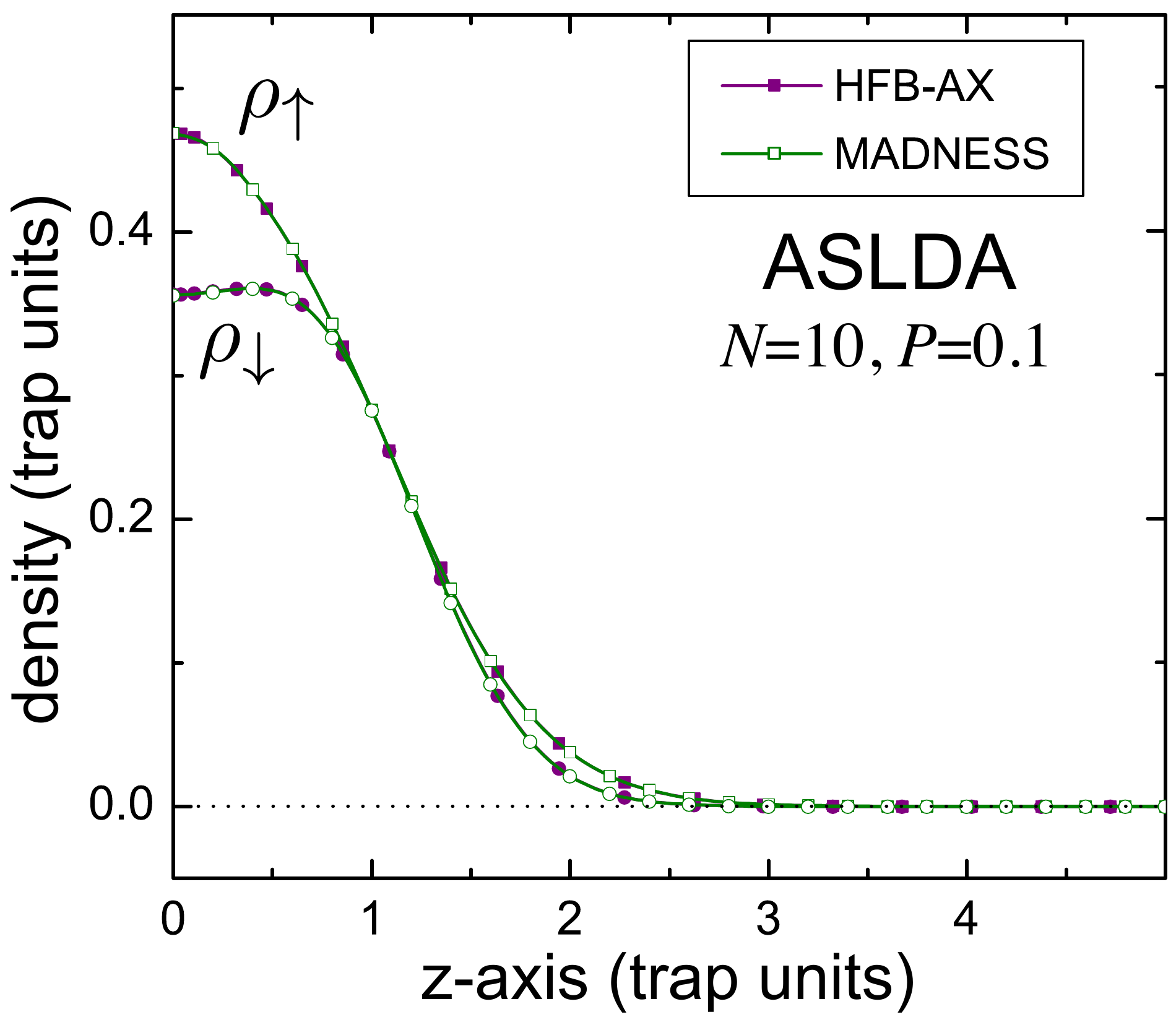}
\caption{\label{den10a} (Color online) Comparison between density distributions
$\rho_\downarrow(\rr)$ and $\rho_\uparrow(\rr)$ obtained in ASLDA with
{\madnesshfb} and {\hfbax}  for  $10$ particles in a
spherical trap with polarization $P=0.1$.}
\label{aslda-densit}
\end{figure}

Next we consider the functional ASLDA, which  was developed to describe
polarized Fermi systems. Because of non-zero spin polarization, the
corresponding HFB solutions break time-reversal symmetry. In the first test, we
performed {\madnesshfb} and {\hfbax} simulations for 10 particles with a total
polarization of $P=0.1$ in a spherical trap. As seen in
Fig.~\ref{aslda-densit}, the density distributions for the spin-up and
spin-down components agree very well between  {\madnesshfb} and {\hfbax}. Some
of the eigenvalues are compared in Table \ref{tabho2}. Note that the
calculation conditions adopted in Table \ref{tabho1} and Table \ref{tabho2} are
the same. It can been seen that the agreement is good up to the 4th digit since
the calculations of local polarization
$x(\rr)=\rho_\downarrow(\rr)/\rho_\uparrow(\rr)$ may lose accuracy in both
approaches when both the spin-up and spin-down densities are very small. In
this case, required precision ALSDA should be significantly greater than that
requested in SLDA calculations.

\begin{table}[htb]
 \caption{\label{tabho2}
Similar as in  Table~\ref{tabho1} but
   for a polarized system in ASLDA.}
\begin{ruledtabular}
\begin{tabular}{ccccc}
  &  \multicolumn{2}{c}{\madnesshfb}  &  \multicolumn{2}{c}{\hfbax} \\
$i$   &  $E_i$  &  $v_i^2$  &  $E_i$  &   $v_i^2$ \\
\hline
\\[-7pt]
1  &  $-0.1333$  &   0.2090    &   $-0.133{\bf 0}$    &  0.209{\bf 1}    \\
2  &  0.0463   &   0.1493    &    0.046{\bf 8}    &  0.149{\bf 4}    \\
3  &  0.0786   &   0.4684    &    0.078{\bf 7}    &  0.468{\bf 2}    \\
4  &  0.8837   &   0.1740(-1)   &    0.883{\bf 8}    &  0.174{\bf 2}(-1)    \\
5  &  1.0157   &   0.2749    &    1.016{\bf 1}    &  0.275{\bf 0}    \\
6  &  1.5425   &   0.2931    &    1.5425    &  0.292{\bf 7}    \\
7  &  1.6944   &   0.3221(-1)   &    1.694{\bf 3}    &  0.322{\bf 5}(-1)    \\
8  &  1.8346   &   0.6160    &    1.834{\bf 8}    &  0.616{\bf 1}   \\
23  &  4.6417  &   0.0155(-1)   &    4.641{\bf 6}    &  0.015{\bf 6}(-1)   \\
24  &  4.8158  &   0.1689(-5)  &    4.815{\bf 7}    &  0.169{\bf 2}(-5)\\
\hline
\\[-7pt]
   &   \multicolumn{2}{c}{$E_t =19.0436$} &  \multicolumn{2}{c}{$E_t=19.04{\bf 43}$} \\
  &  \multicolumn{2}{c}{ $(\lambda_\uparrow+ \lambda_\downarrow)/2=2.1684 $}  &  \multicolumn{2}{c}{$(\lambda_\uparrow+ \lambda_\downarrow)/2=2.168{\bf 3}$} \\
  &  \multicolumn{2}{c}{ $N_\uparrow-N_\downarrow=1.0034 $}  &  \multicolumn{2}{c}{$N_\uparrow-N_\downarrow=1.003{\bf 38}$}
\end{tabular}
\end{ruledtabular}
\end{table}

\begin{figure}[t]
\includegraphics[width=0.85\columnwidth]{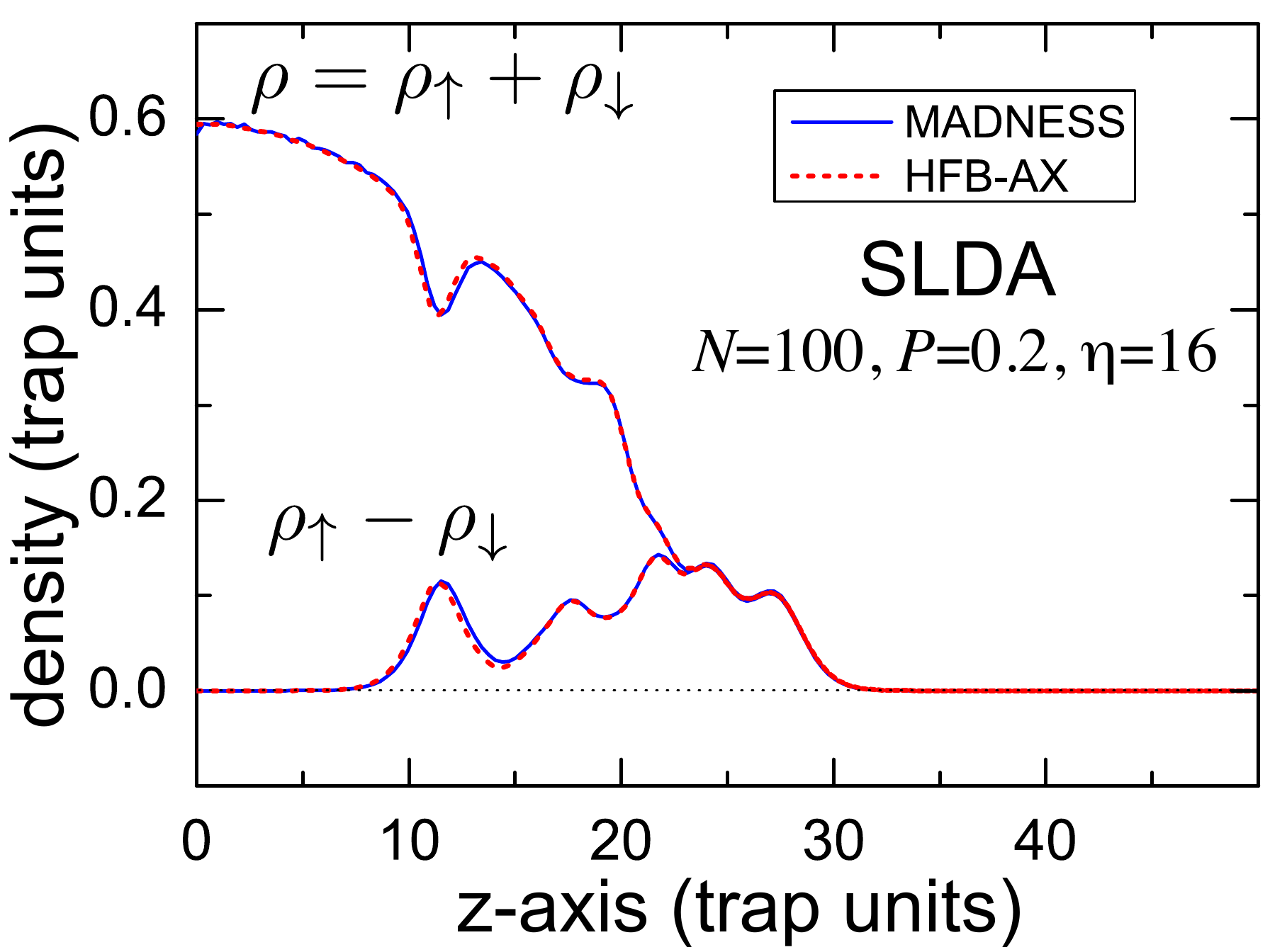}
\caption{ \label{slda16d} (Color online)
Comparison between density distributions
$\rho_\uparrow + \rho_\downarrow$ and $\rho_\uparrow - \rho_\downarrow$
 obtained in SLDA with
{\madnesshfb} and {\hfbax}
for 100 particles with $P=0.2$ in an elongated trap with $\eta=16$.
}
\end{figure}

\begin{figure}[t]
\includegraphics[width=0.85\columnwidth]{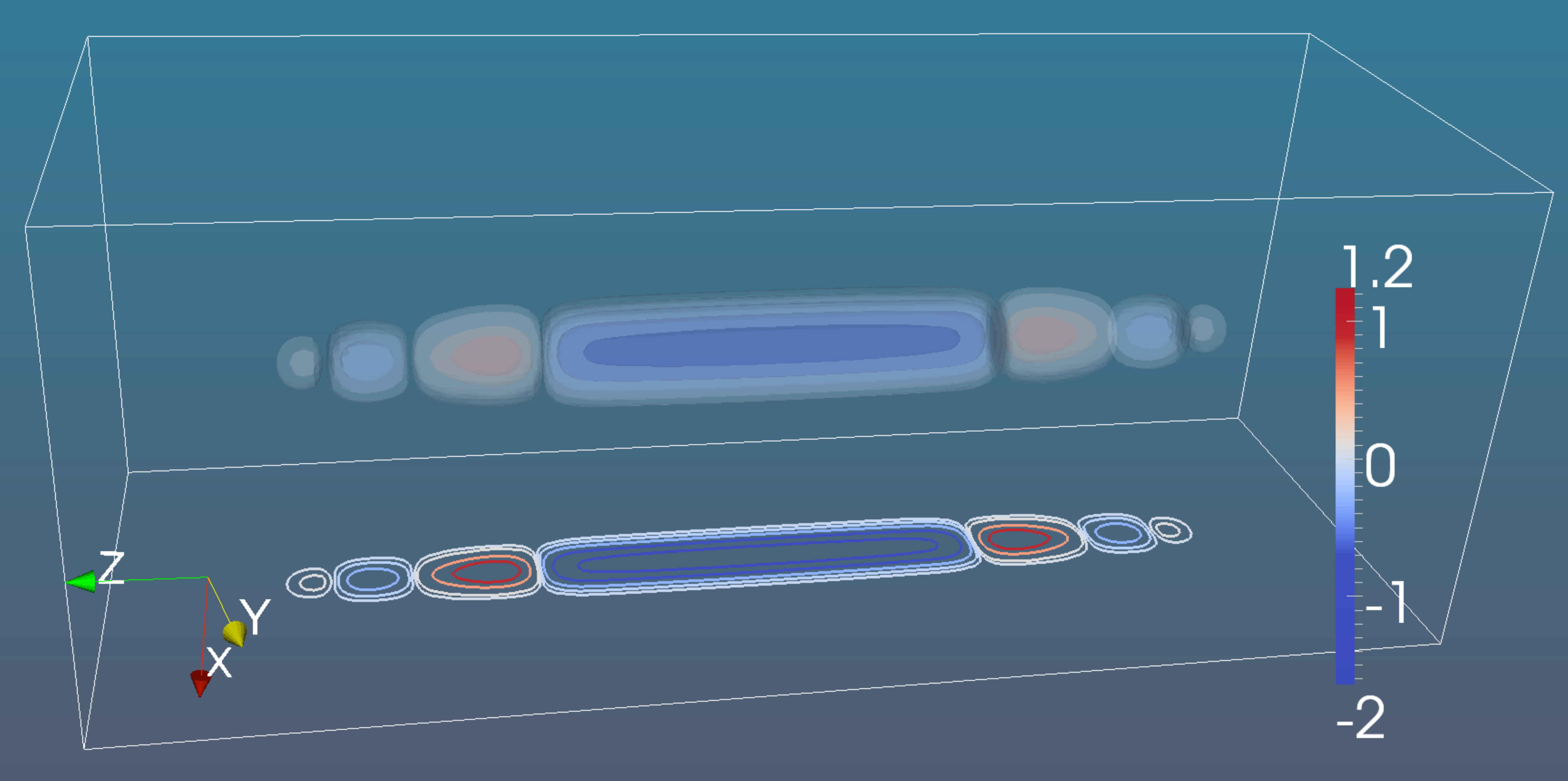}
\caption{\label{slda16p} (Color online)
The pairing potential of 100 particles with $P=0.2$ in an elongated trap with $\eta=16$ computed with SLDA.
}
\end{figure}

To demonstrate the capability of {\madnesshfb} for accurate simulation of large
systems, we carried out SLDA simulations for 100 particles with polarization
$P=0.2$ in an elongated trap with $\eta=16$. The choice of SLDA was motivated
by the above-mentioned loss accuracy of ASLDA caused by a numerical error on
$x(\rr)$ at low densities (large distances). The simulation box is $(-L,L)^3$
with $L=120$. This computation involves about 2,000 eigenstates and 5,000 cores
on Titan supercomputer, and takes about 4 hours to reach convergence. The total and polarization densities for the {\madnesshfb}
and {\hfbax} simulations are shown in Fig.~\ref{slda16d}. The 3D pairing
potential is displayed in Fig.~\ref{slda16p}. The oscillations of the pairing
field in a spin-polarized system, characteristic of the Larkin-Ovchinnikov
phase, are clearly seen (see Ref.~\cite{(Pei10)} for more discussion).


\subsection{Skyrme HF+BCS solver for nuclei}\label{HFBCS}

Most of the currently-envisioned applications of {\madnesshfb} pertain to the
nuclear many-body problem. To this end, the adaptive multiresolution Skyrme-HFB
solver has also been developed. The {\madnesshfb} approach for nuclei is
similar to the SLDA for cold atoms but much more involved due to the continuum
discretization, as the atomic nucleus is an open system and associated boxes
are large \cite{(Pei11)}. Therefore, as an initial step, we carry out Skyrme HF
and Skyrme HF+BCS  calculations and benchmark them with HFODD.

For both HF and HF+BCS calculations, we consider the neutron-rich nucleus
$^{110}$Mo which is triaxially deformed in its ground state in some models
\cite{(Shi13)}. We use SkM*\,\cite{(Bar82)} Skyrme parametrization, and take
$\hbar^2$/2m=20.73\,MeV\,fm$^2$ for benchmarking purpose.

In  pairing calculations for $^{110}$Mo, due to the small neutron separation
energy, the positive-energy HF levels are important as they participate in
pairing. This creates a problem when trying to compare BCS or HFB  results
based on solvers using  coordinate-space framework and harmonic-oscillator
expansion as the continuum representation is different in both approaches.
Indeed, coordinate-space solvers {\madnesshf} or {\madnesshfbcs}, when applied
to large boxes, produce a very dense unbound single-neutron spectrum
\cite{(Dob96),(Bor06),(Pei11)}. On the other hand, the single-neutron spectrum
of oscillator-based  {\hfodd} is fairly sparse. Therefore, to minimize the
difference between these two codes for a meaningful benchmarking, we switch off
neutron pairing and retain only bound 70 single-proton orbits in the BCS phase
space. We adopt mixed density dependent delta interaction\,\cite{(Dob02c)}. The
proton pairing strength is chosen to be $-$500\,MeV to obtain paired solution.
Our {\madnesshf} and {\madnesshfbcs} calculations are performed in a  large 3D
box $(-L,L)^3$ with $L=50$\,fm. The wavelet order is $p=9$ with requested truncation
precision $\epsilon=10^{-7}$. {\hfodd} calculations are performed with 1140 and 1540
spherical harmonic oscillator states, corresponding to 17 and 19 shells,
respectively. The oscillator constant is 0.4975890\,fm$^{-1}$.

Since MADNESS calculations are numerically extensive, it is desirable to
warm-start the runs with wave functions (or densities) from the converged
{\hfodd} solution. We have implemented such an interface between {\hfodd} and
{\madnesshfb}.

Table\,\ref{tabho4} compares {\madnesshf} and {\hfodd} results for the  triaxial
ground-state configuration in  $^{110}$Mo.
\begin{table}[htb]
\caption{\label{tabho4}
Comparison between results of {\madnesshf} and {\hfodd} for  the triaxial
ground state of $^{110}$Mo: total binding energy $E_t$,  kinetic
energy, $E_{\rm kin}$, Coulomb energy  $E_c$, and spin-orbit energy $E_{\rm SO}$ (all in MeV),
mass r.m.s. radius $R_{\rm rms}$ (in fm), and mass quadrupole  moments $Q_{20}$ and
 $Q_{22}$ (in fm$^2$).
The ``0-th iter" column shows {\madnesshf} warm-start numbers at the beginning of the iteration process  with wave functions
and densities imported from   converged {\hfodd} results using 1140 basis states.
}
\begin{ruledtabular}
\begin{tabular}{lrrrr}
  & {\hfodd}  & {\hfodd} & {\madnesshf}  &  {\madnesshf} \\
  &  (1140) & (1540) & (0-th iter)   &  (converged) \\
\hline
\\[-7pt]
$E_t $       & $-$921.803   & $-$921.932 &  $-$921.808  &  $-$922.119\\
$E_{\rm kin}$&  1998.074  & 1998.316 & 1998.075   &  1998.846 \\
$E_{c}$      &  251.116   & 251.128  & 251.116    &  251.138 \\
$E_{\rm SO}$ &  $-$69.290 & $-$69.273 & $-$69.290 & $-$69.276 \\
$R_{\rm rms}$&  4.6696    & 4.6697   & 4.6696     &  4.6697 \\
$Q_{20}$     &  914.12    & 913.58   & 914.12     & 913.69 \\
$|Q_{22}|$   &  367.93    & 368.48   & 367.93     & 368.88
\end{tabular}
\end{ruledtabular}
\end{table}

\begin{figure}[hbt]
\includegraphics[width=0.9\columnwidth]{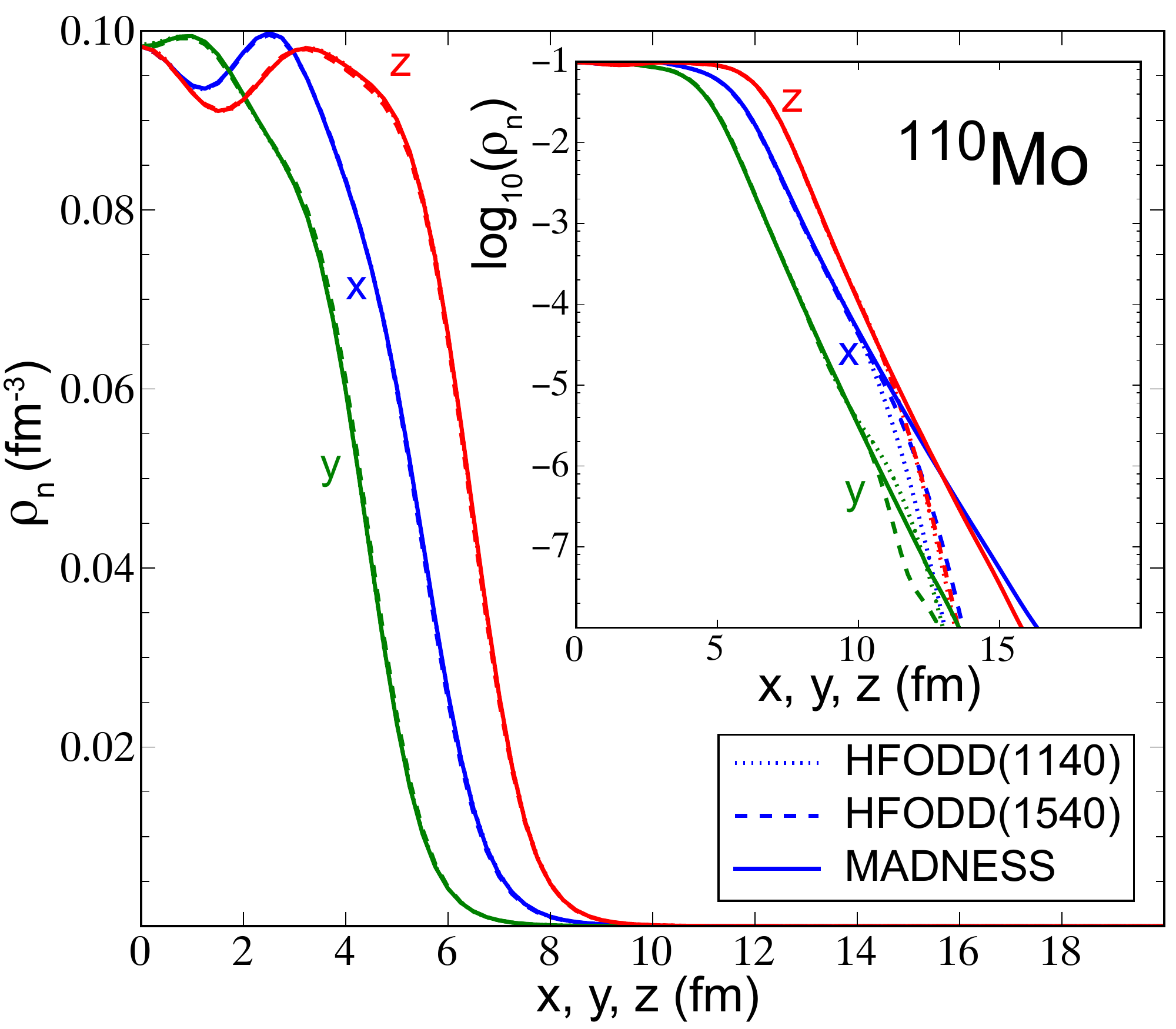}
\caption{\label{den-mo-hf} (Color online) Neutron density distribution for $^{110}$Mo
in {\madnesshf} (solid line), {\hfodd}(1140) (dotted line), and {\hfodd}(1540) (dashed line)
along $x$-, $y$-, and $z$-axis, moving from the origin.
The inset (in a logarithmic scale) illustrates the tail behavior of density.}
\end{figure}

The {\madnesshf} results labeled ``0-th iter" are warm-start initialization
numbers, with densities imported from {\hfodd}(1140). As expected, ``0-th iter"
and {\hfodd}(1140) values are extremely close. A very small difference
$\approx$5\,eV on the total energy can be attributed to the potential (Skyrme)
energy. In particular, the density-dependent term ($\sim\rho^{\gamma+2}$)
produces the largest difference. The excellent agreement between these two
calculations indicates that the interface between the two solvers has been
implemented correctly, and that the individual Skyrme EDF terms have been coded
properly in {\madnesshf}. By increasing the basis size in {\hfodd} to 1540
states, the total binding energy decreases by $\sim$130\,keV. However, it is
still $\sim$190\,keV above the {\madnesshf} result. This difference can be
traced back to asymptotic behavior of nucleonic densities obtained in the two
solvers. Figure.\,\ref{den-mo-hf} displays the neutron density profiles along
$x$-, $y$-, and $z$-axis (moving from the origin) computed in {\hfodd}(1140),
{\hfodd}(1540), and {\madnesshf}. When displayed in linear scale, one can
hardly see a difference between {\hfodd} and {\madnesshf} predictions. However,
when inspecting the density in a logarithmic scale, one can see a
characteristic damping at large distances (10-12\,fm) in {\hfodd} due to the
finite size of oscillator basis. We recall that the {\madnesshf} calculations
were carried out in a box extending to 50\,fm.

Finally, Table~\ref{tabho5} displays  HF+BCS
results. Again, the agreement between {\madnesshfb}  and {\hfodd} is excellent, with the total binding energy in {\madnesshfb}  being $\sim$150\,keV below that of  {\hfodd}(1540).
\begin{table}[htb]
\caption{\label{tabho5}
Similar to table\,\ref{tabho4}, except that we include BCS pairing
for protons, see text for details.
  }
\begin{ruledtabular}
\begin{tabular}{lrrrr}
  & {\hfodd}  & {\hfodd} & {\madnesshf}  &  {\madnesshf} \\
  &  (1140) & (1540) & (0-th iter)   &  (converged) \\
\hline
\\[-7pt]
$E_t $       & $-$922.419 & $-$922.549  & $-$922.425  & $-$922.707  \\
$E_{\rm pair}$    & $-$4.981 & $-$4.988    & $-$4.981  & $-$4.781  \\
$\lambda_p$  & $-$12.688 & $-$12.692   & $-$12.688  & $-$12.697  \\
$E_{\rm kin}$& 1998.055 &  1998.285   & 1998.055  & 1998.607  \\
$E_{c}$      & 251.239 & 251.252   & 251.239  & 251.250  \\
$E_{\rm SO}$ & $-$67.251 & $-$67.228  & $-$67.251  & $-$67.220 \\
$R_{\rm rms}$& 4.6610 & 4.6611  & 4.6610  & 4.6615  \\
$Q_{20}$     & 859.64 &  858.74   & 859.64  & 860.91 \\
$|Q_{22}|$   & 355.92 & 356.58 & 355.92  & 357.91
\end{tabular}
\end{ruledtabular}
\end{table}

\section{Summary}\label{conclusions}

In this paper, we introduce nuclear DFT framework based on the adaptive
multi-resolution 3D HFB solver {\madnesshfb}. The numerical method employs
harmonic analysis techniques with multiwavelet basis; user-defined finite
precision is guaranteed. The solver applies state-of-the-art in parallel
programming techniques that can take advantage of high performance
supercomputers.

Applications have been presented for polarized ultracold atoms in very
elongated traps and for triaxial neutron-rich nuclei. The solver has been
benchmarked against other advanced HFB solvers: a 2D coordinate-space solver
{\hfbax} based on B-spline technique and a 3D solver {\hfodd} employing the
harmonic oscillator basis expansion. The advantage of {\madnesshfb} is its
ability to treat large and complex systems without restriction on symmetries.
Examples of future nuclear structure applications include: weakly bound nuclei
with large spatial extensions, heavy-ion fusion, nuclear fission, complex
topologies in super- and hyperheavy nuclei \cite{(Naz02),(Dec03),(Jac11)}, and
pasta phases in the inner crust of neutron stars
\cite{(Rav83),(New09),(Dor12),(Sch13)}. Future atomic applications of
{\madnesshfb} include description of large number of fermions ($\sim10^5$) in
highly elongated optical traps ($\eta \sim 50$) \cite{(Par06)}.

\begin{acknowledgments}
Useful discussions with N. Hinohara, J. Sheikh, and N. Schunck are gratefully 
acknowledged.
This work was supported by the U.S. Department of
Energy (DOE) under Contracts No. DE-AC05-00OR22725 (ORNL), No.
DE-FG02-96ER40963
(University of Tennessee),  No.
DE-SC0008499    (NUCLEI SciDAC Collaboration), No. DE-FG02-13ER42025
 (China-U.S. Theory Institute for Physics with Exotic Nuclei);
by the National Natural Science Foundation of China under Grant No.11375016, 11235001.
An award of computer
time was provided by the National Institute for Computational
Sciences (NICS) and the Innovative and Novel Computational Impact on
Theory and Experiment (INCITE) program using resources of the OLCF and ALCF facilities.
\end{acknowledgments}

\bibliographystyle{apsrev4-1}
\bibliography{madness}

\end{document}